\documentstyle[12pt]{article}
\newcommand\bm[1]{\mbox{\boldmath$#1$}}
\begin{document}
\title{On the construction of global models describing 
rotating bodies; uniqueness of the
exterior gravitational field}
\author{Marc Mars$^\natural$$^\ast$ and Jos\'e M. M.
Senovilla$^\flat$\thanks{Also at Laboratori de F\'{\i}sica Matem\`atica,
Societat Catalana de F\'{\i}sica, IEC, Barcelona.} \\
$\natural$ Institute of Theoretical Physics, University of Vienna \\
Boltzmanngasse 5, Wien A-1090, Austria. \\
$\flat$ Departament de F\'{\i}sica Fonamental, Universitat de Barcelona\\
 Diagonal 647, 08028 Barcelona, Spain.}
\maketitle
\begin{abstract}
The problem of constructing global models describing isolated
axially symmetric rotating bodies in equilibrium is analyzed. It is
claimed that, whenever the global spacetime is constructed by giving boundary
data on the limiting surface of the body and integrating Einstein's
equations both inside and ouside the body, the problem becomes overdetermined. 
Similarly, when the spacetime describing the {\it interior} of the body is
explicitly given,
the problem of finding the {\it exterior} vacuum solution becomes
overdetermined.
We discuss in detail the
procedure to be followed in order to construct the exterior vacuum field
created by a given but arbitrary distribution of matter.
Finally, the uniqueness of the exterior vacuum gravitational field
is proven by exploiting the harmonic map formulation of the vacuum equations
and the boundary conditions prescribed from the matching. 
\end{abstract}

\vspace{1cm}

The description and analysis of isolated 
rotating objects in equilibrium is a fundamental problem in general
relativity. Unfortunately, despite important efforts in that direction,
our present understanding of the question
is still very incomplete. In particular, not a single model describing both
the interior and the exterior of a selfgravitating, spatially compact,
rotating object is explicitly known. There are, however, very interesting
solutions which describe the exterior
field of a physical object. We are referring
to the solution for the rotating disc of dust recently found by Neugebauer
and Mainel \cite{Neu}. This metric is vacuum everywhere
and contains discontinuities in the first derivatives of the metric
across an equatorial disc centered on the axis of symmetry, which
can be interpreted as a shell of dust.
Despite its manifest interest and physical relevance,
this solution does not contain an interior body in strict
sense (in mathematical
terms, the interior of the body is empty). In general, solving the
vacuum field equations everywhere and allowing for discontinuities in the
first derivatives of the metric will be adequate for describing
the gravitational field generated by a shell of matter, or as a possibly
good approximation of some very thin objects, but it cannot be used as an
exact solution when the source object has a non-empty interior. In that case,
the problem becomes even more difficult because the global model must be
constructed from two spacetimes, one describing the interior
of the body (i.e. a solution of Einstein's field equations with matter) and
another one describing the exterior field (i.e. an asymptotically flat
solution of the vacuum equations). Several methods have been
presented recently for attacking this kind of problem. One of them 
requires the resolution of Einstein's field equations both
inside and outside the body starting from a common boundary.
This approach is taken, for instance, in a recent paper \cite{prl} where the
existence, regularity and uniqueness
of the solution has been investigated.
More precisely, the authors prescribe
Dirichlet boundary data on a two-sphere and then show that both the
vacuum field equations and the interior problem (assuming
rigid rotation) are solvable. 
The interest of these results is obvious, but it must be noted that
they still {\it do not} prove the solvability of the global problem,
not even for the rigidly rotating case. The
reason is that it cannot be ensured that the two solutions thus
constructed match appropriately across the boundary hypersurface.
Given the type of boundary data used in \cite{prl}, it follows
that the global spacetime has indeed a continuous metric, but
nothing can be said a priori about the first derivatives of the metric (they
can certainly be discontinuous). Hence, in general, the global spacetime will
have an undesired and uncontrolled surface layer of matter, which
is an important weakness of the method. The same argument shows in
general that,
whenever two different spacetime pieces (interior and exterior) are used to
construct the global manifold, the physical problem requires imposing
the {\it full} set of matching conditions and not just a (mathematically
suitable) subset of them.

In order to solve the global problem one can alternatively adopt the
point of view of finding the whole spacetime once and for all.
The most favourable situation would be finding the equivalent of a Green
function for the coupled system of equations describing gravity and matter,
so that the solution can be expressed directly in terms of the sources. 
Although this is probably the mathematically most well-motivated method,
the technical difficulties are enormous and very scarce results are
known at present. In particular, it seems very unlikely that an explicit
model can be obtained using this method in the short, or even in the
medium, term. It must be mentioned, however, that this approach has been
followed in a very interesting paper by Heilig
\cite{U}, where the existence of global solutions describing rigidly rotating
stars is proven for configurations sufficiently close to a Newtonian solution.
A second global method that can be employed requires
finding smooth solutions of the differential system
describing gravity and matter, such that the matter fields
tend to zero fast enough when we approach infinity
(so that the solution can still be interpreted as an isolated body).
Although this method is potentially capable of producing explicit
solutions, its main drawback is that the rotating body is not truly 
finite in size because it has no limiting surface.
Hence, the solutions have no vacuum exterior.
Of course this does not mean that the solutions are necessarily unphysical, but
for many purposes it is highly convenient to have models in which the
rotating object has a clear limiting boundary. A second disadvantage shared
by the two global methods described above 
is that the system of differential equations to be analysed
must be closed, and this requires the imposition of conditions on the type of
matter beforehand such as, for instance,
rigid rotation and a barotropic equation of state. This
means that a different analysis must be performed for each different type of
matter.

All in all, it follows that for any truly isolated and finite
body, the full set of matching conditions between the interior
metric and the exterior vacuum solution must be inevitably imposed in order to
analyse the physical problem. Excluding the ``Green function'' approach, 
the construction of the global solution requires three steps, namely, 
obtaining the interior metric (either by solving the Einstein field
equations for an adequate type of matter or by any other means),
solving the exterior vacuum field equations and imposing the matching
conditions between the two spacetimes.
These three steps could be solved, in principle, in any order. In our opinion,
however, the natural ordering is to provide first the interior metric
and then finding the exterior metric which satisfies the matching
conditions with the prescribed interior region. This point of view
incorporates in a natural way the common idea that the exterior field
is created by the interior distribution of matter. Moreover, the natural
questions of existence and uniqueness of the gravitational field are
only reasonable for the exterior problem, once the interior metric is known.
The reason is that we cannot expect existence, or even uniqueness, of
the interior region for a given exterior gravitational field
(it is well-known that in Newtonian gravity different distributions
of matter produce the same exterior gravitational field; also, the spherically
symmetric case proves that this still holds in General Relativity).
Moreover, the inside-outside procedure has the important
advantage that the resolution of the interior
field equations, which are much more difficult to handle that the
vacuum equations, can be performed by any method we wish, or even not done
at all if we prescribe the interior metric by hand.
In other words, there is no need to solve the interior equations
in order to study the associated exterior problem. Obviously, for physically
relevant problems the interior field equations should be
solved, but the important point here is that the interior problem
is completely decoupled from the exterior one.

The aim of this letter is, besides stressing and clarifying
all the points above, to describe
in detail the conditions one obtains by matching an
interior (arbitrary but supposed to be known) stationary and axially
symmetric metric with an unknown exterior vacuum metric, as well as to analyze
and solve some of the problems and difficulties that arise.
In particular, we show that {\it the junction
conditions introduce two new essential parameters in the interior}
metric. To the best of our knowledge, this fact was unknown hitherto. We
also show that, for each value of these two parameters,
the matching hypersurface, both from the exterior and the interior
regions, is  uniquely determined. Furthermore, the matching conditions
fix the boundary conditions for the exterior problem. When written in terms
of the Ernst potential \cite{Ernst}, the boundary value
exterior problem turns out to be {\it overdetermined} and {\it not unique}.
It is overdetermined because the
interior geometry of the self-gravitating object provides
more boundary conditions than necessary for solving the Ernst equations
and it is not unique for, as we shall see, the boundary conditions fix the
Ernst potential on the matching hypersurface up to an arbitrary, non-trivial,
additive constant. Since the Einstein field
equations for stationary fields are of elliptic type,
it follows that the physical problem is
overdetermined and mathematically not well-posed.
%These are additional difficulties in treating the subject but they cannot
%be avoided if the physical problem is to be treated: they must be
%addressed and,if possible, solved.
After computing and classifying the matching conditions, we finally prove
the {\it uniqueness} of the exterior vacuum gravitational field generated by a
given isolated, axially symmetric, rotating object in equilibrium. Although
this result is quite natural and it has been often assumed implicitly
in numerical calculations, no proof is available in the literature.
The only results we are aware of 
are due to  Lichnerowicz \cite{L}, but they only apply doubly locally,
in the sense that they hold only in a  neighbourhood of a  local
piece of the matching hypersurface. 
The method we use for proving the uniqueness rely on the harmonic
map formulation of the exterior vacuum field equations. This provides
a proof which is extremely simple and which is
valid for {\it any} boundary data (by using more
classical methods of functional analysis one normally obtains theorems
which only hold for boundary data satisfying certain inequalities).
It must be emphasized here that none of the recent 
results on the uniqueness of solutions of the Ernst
equations -- see e.g. \cite{We} and
references therein --, analyzes the problem we are
considering in this letter. 

Let us then describe in detail the procedure discussed above
for constructing global
models for rotating objects. Let us assume we are given a
stationary and axisymmetric spacetime $(V_I, g_I)$ 
describing the interior of the
selfgravitating fluid. We make the usual assumption that this metric satisfies
the circularity condition \cite{Pap} (for fluids without energy flux,
this is equivalent \cite{Car} to the absence of convective motions).
Hence, there exist coordinates $\{T,\Phi,r,\zeta\}$
in which the line-element reads
\begin{eqnarray}
ds^2_I= - e^{2V} \left ( dT + B d \Phi \right)^2 + e^{-2V} \left [
e^{2h} \left (dr^2 + d\zeta^2 \right) + \alpha^2 d \Phi^2 \right ]
\label{intmet}
\end{eqnarray}
where $V$, $B$, $h$ and $\alpha$ are (known) functions of $r$ and $\zeta$
and $\partial_{\Phi}$ is the axial Killing.
The searched vacuum exterior manifold $(V_E, g_E)$ is assumed truly
stationary (free of ergoregions and/or Killing horizons) so that
it can be globally described using Weyl coordinates
\begin{eqnarray}
ds^2_E = -e^{2U} \left ( dt + A d \phi \right )^2 + e^{-2U} \left [
e^{2k} \left ( d \rho^2 + dz^2 \right )+ \rho^2 d\phi^2 \right ]
\label{Vacon}
\end{eqnarray}
where $U$, $A$ and $k$ are functions of $\rho$ and $z$ only. The
axial Killing vector of this spacetime is
$\partial_\phi$ and the axis of symmetry is located at $\rho=0$. The
coordinates have been chosen so that 
$\partial_t$ is the unique timelike Killing vector which is unit
at infinity (i.e. $t$ mesures  proper time at infinity)
Then, the coordinate freedom in (\ref{Vacon}) consists only of
trivial constant shifts of $t$, $\phi$ and $z$.
Introducing the Ernst potential \cite{Ernst}
${\cal E} \equiv e^{2U} + i\, \Omega$, where $\Omega$ is defined up to
an additive constant by
\begin{equation}
\rho\, \Omega_{,\rho} = - e^{4U} A_{,z}\, , \hspace{1cm} \rho\, \Omega_{,z} =
e^{4U} A_{,\rho}\, , \label{defOme}
\end{equation}
(a comma indicates partial derivative), the vacuum field equations read
simply
\begin{equation}
2\, \triangle U + e^{-4U} \langle d \Omega, d \Omega \rangle = 0\, ,
\hspace{1.5cm} \triangle \Omega - 4  \langle d \Omega, d U
\rangle  = 0\, , \label {EQ}
\end{equation}
where $d$ is the exterior derivative, $\triangle$ is the
laplacian of the 3-Euclidean metric 
\begin{eqnarray*}
ds_{M}^2 = d\rho^2 + dz^2 + \rho^2 d\phi^2
\end{eqnarray*}
and $\langle\, ,\, \rangle$ denotes scalar product in this flat
space. The equation for the remaining function $k$ is
a simple quadrature integrable once the solution of (\ref{EQ})
is obtained. Thus, the system of differential equations to be
solved constitutes a coupled, non-linear, quasi-linear elliptic system of
p.d.e.
where the principal part is the Laplace-Beltrami operator on the
Euclidean 3-space.

In order to find the boundary conditions of this problem,
the matching of the known interior $\left
(V_I, g_I \right)$ with the undetermined exterior $\left(V_E, g_E \right)$ must
be analyzed. The standard junction theory
involves finding two isometric imbeddings of a 3-dimensional
manifold $(\Sigma, g_\Sigma)$ into $(V_I,g_I)$ and $(V_E,g_E)$
respectively such that the two second fundamental forms in $\Sigma$
inherited from both imbeddings coincide (see e.g.\ \cite{MS}).
It is natural to assume that $\Sigma$ preserves both the stationarity
and the axisymmetry and hence
local coordinates  $\{\lambda, \tau, \varphi \}$
in $\Sigma$ can be chosen so that the imbedding
$\chi_E\,:\,\Sigma \longrightarrow V_E$ reads
\begin{equation}
\chi_E : \left \{t = \tau, \,\,\phi = \varphi, \,\,\rho = \rho (\lambda), \,\,
z = z(\lambda) \right \}, \label{embext} 
\end{equation}
that is, we adapt the coordinates in $\Sigma$ so that $\partial_\tau$ and 
$\partial_{\varphi}$ coincide with $\partial_t$ and $\partial_\phi$ on
the exterior matching hypersurface $\chi_E(\Sigma)$. The next task is the
determination of the imbedding $\chi_I\, :\, \Sigma \longrightarrow V_I$.
Obviously, the images of
$\partial_\tau$ and $\partial_\varphi$ in $\chi_I \left(\Sigma \right)$ must be
Killing vectors.
Since the axial Killing vector has an intrinsic definition as
the only one with $2\pi$-periodic closed orbits and a symmetry axis,
we must impose 
$\chi_{I*} \left (\left . \partial_\varphi \right |_\Sigma \right ) =
\left . \partial_\Phi \right |_{\chi_I(\Sigma)}$
($\chi_{I*}$ is the push-forward of $\chi_I$).
Regarding the image of $\partial_\tau$, and given that
the stationary Killing vector in $\left(V_I, g_I \right)$ does not have
an intrinsic characterization, we must allow for
\vspace{-8mm}
\begin{eqnarray}
\chi_{I*} \left (\left . \partial_\tau \right |_{\Sigma} \right ) =
a \left . \left (   \partial_T +
b \partial_\Phi \right ) \right |_{\chi_I(\Sigma)}, \label{matab}
\end{eqnarray}
where $a$ is a positive constant (to preserve the time
orientation) and $b$ is a constant restricted only
to preserve the timelike character of the Killing.
In order to deal with this freedom, we perform in $\left(V_I, g_I \right)$
the coordinate change
\vspace{-2mm}
\begin{eqnarray}
\Phi = \Phi^\prime + a b T^\prime, \hspace{1cm} T = a T^\prime
\label{newco}
\end{eqnarray}
which implies
$\partial_{T^\prime} = a \left (\partial_T + b \partial_\Phi \right)$ and
$\partial_{\Phi^{\prime}} = \partial_{\Phi}$, leaving the axial Killing
invariant.
The new metric potentials (defined so that (\ref{intmet}) keeps the same
form when all unprimed quantities are substituted by their primed
counterparts) are $\alpha^\prime = a \alpha ,\, h^\prime -V'= h-V$,
\begin{eqnarray*}
e^{2 V^\prime} = a^2 \left [ \left (1+ b B \right)^2 e^{2V} - 
\alpha^2 b^2 e^{-2V} \right ], \hspace{1cm}
B^\prime =  \frac{ B \left (1+ b B \right) e^{2V} - \alpha^2
b e^{-2V}}{a \left (1 +b B \right)^2 e^{2V} - a \alpha^2 b^2 e^{-2V}}\, .
\end{eqnarray*}
Thus, the matching procedure has introduced two new {\it essential}
parameters
in the interior metric (even though they are irrelevant for the interior
geometry by itself). In order to clarify this, let us show that matchings with
different values of $a$ and $b$ are essentially inequivalent from the exterior
point of view, by taking as an example an interior describing a non-convective
fluid with velocity vector
$\vec{u} = N \left (\partial_T + w \partial_{\Phi} \right)$,
where $N$ and $\omega$ are functions of $r$ and $\zeta$. For any matching
satisfying (\ref{matab}), $\vec{u}$ on the matching hypersurface
as seen from the exterior becomes
\begin{eqnarray}
\left . \vec{u} \right |_{\Sigma} = \left .N \left [ \frac{1}{a} \partial_t
+ \left (w-b \right ) \partial_\phi \right ] \right |_{\Sigma} \label{velext}
\end{eqnarray}
(we will identify $\Sigma\equiv \chi_I \left (\Sigma \right)\equiv \chi_{E}
\left (\Sigma \right)$ from now on to simplify notation).
The coordinates $t$ and $\phi$ in the asymptotically flat exterior have an
intrinsic meaning and (\ref{velext}) shows that the proper angular velocity of
the fluid on $\Sigma$ is $a \left (w -b \right)$, which depends essentially on
$a$ and $b$. Thus, different values of $a$ and $b$ will
potentially give rise to different exteriors (whenever they exist), because the
state of motion of the interior body is distinct as seen from infinity.
Consequently, we should not expect that the exterior gravitational field for
different values of $a$ and $b$ coincides, and the physically relevant
problem is the uniqueness of the exterior gravitational field when $a$ and $b$
are fixed. In other words, the exterior gravitational field will be fixed when
the interior metric {\it and} the identification of the interior with the
exterior through $\Sigma$ are both prescribed. 

The new coordinate system (\ref{newco}) will be used from now on dropping
primes everywhere. The imbedding $\chi_I\, :\, \Sigma \longrightarrow V_I$
reads then
\begin{eqnarray}
\chi_I : \left \{   T = \tau, \Phi = \varphi,  r = r (\lambda), 
\zeta = \zeta(\lambda)  \right \} . \label{embint}
\end{eqnarray}
The four unknowns $\rho(\lambda)$, $z(\lambda)$ in (\ref{embext})
and $r(\lambda)$, $\zeta(\lambda)$ in (\ref{embint}) define the junction
hypersurfaces and must be determined from the matching conditions. It can be
shown \cite{Mer} that all the matching equations
can be reorganized into the following three sets of conditions (to prove the
equivalence the vacuum field equations for the exterior have to be used):

\noindent
{\bf 1} {\it Conditions on the interior hypersurface.}
The functions $r\left(\lambda\right)$, $\zeta \left (\lambda \right)$ defining
the interior hypersurface are the solutions of the overdetermined
system of ordinary differential equations
\begin{eqnarray}
\left .G_{22} \dot{\zeta} - G_{23} \dot{r} \right |_{\Sigma} = 0, \hspace{1cm} 
\left .G_{32} \dot{\zeta} - G_{33} \dot{r} \right |_{\Sigma} = 0, \label{matc}
\hspace{8mm}
\end{eqnarray}
where the $G$'s are the components of the Einstein tensor of
(\ref{intmet}) in the orthonormal tetrad
\begin{eqnarray*}
\bm{\theta^0} = e^{V} \left (d T + B d \Phi \right), \hspace{9mm}
\bm{\theta^1} = \alpha e^{- V} d \Phi, \hspace{9mm}
\bm{\theta^2} = e^{h- V} dr, \hspace{9mm}
\bm{\theta^3} = e^{h-V} d\zeta,
\end{eqnarray*}
and the dot denotes derivative with respect to $\lambda$.
In  many cases, these two differential equations are incompatible and no
matching hypersurface exists, meaning that the given metric cannot be considered
as a model for describing an isolated body with a vacuum exterior.
There are also particular energy-momentum tensors for which the above equations
do not provide any information at all (those with $G_{22}=G_{23}=G_{33}=0$,
including the important case of dust), so that the matching can be performed, in
principle, at any timelike hypersurface preserving the symmetry. However,
as is clear from (\ref{matc}), the interior matching hypersurface is unique
in the generic case.

\noindent
{\bf 2} {\it Exterior matching hypersurface}.\
The exterior matching hypersurface $\rho(\lambda),z(\lambda)$ is then uniquely
determined by
\begin{eqnarray}
\rho(\lambda) = \left . \alpha \right |_\Sigma \, , \hspace{1cm}
\dot{z}\left(\lambda\right) =  \left . \alpha_{,r} \dot{\zeta} -
\alpha_{,\zeta} \dot{r} \right |_\Sigma \label{BS}
\end{eqnarray}
(the additive constant in $z\left(\lambda \right)$ is spurious given the
shift freedom $z \longrightarrow z + \mbox{const.}$).

\noindent
{\bf 3} {\it Boundary conditions for the exterior problem.}
Finally, the matching provides boundary conditions on
the exterior metric potentials $U$ and $A$ at the matching hypersurface
\begin{eqnarray}
\left . U \right | _\Sigma = \left . V \right |_\Sigma, \hspace{1cm}
\left . A \right |_\Sigma = \left . B \right |_\Sigma, \hspace{35mm}
\nonumber \\
\left . U_{,\rho} \dot{z} -  U_{,z} \dot{\rho} \right |_\Sigma =
\left . V_{,r} \dot{\zeta} - V_{,\zeta} \dot{r} \right |_\Sigma , \hspace{1cm}
\left . A_{,\rho} \dot{z} -  A_{,z} \dot{\rho} \right |_\Sigma =
\left . B_{,r} \dot{\zeta} - B_{,\zeta} \dot{r} \right |_\Sigma.
\label{BC}
\end{eqnarray}
>From these equations we must get the appropriate boundary conditions for the
exterior vacuum problem in terms of the Ernst potential. This 
can be done by noticing that 
\begin{eqnarray*}
\left . \dot{ \Omega} \right |_\Sigma = \left .
\frac{}{} \Omega_{,\rho} \dot{\rho} + 
\Omega_{,z} \dot{z} \right |_{\Sigma} = 
\left . \frac{e^{4U}}{\rho} \left ( - A_{,z} \dot{\rho} + A_{,\rho}
\dot{z} \right ) \right |_{\Sigma} = \left .\frac{e^{4V}}{\alpha}
\left (  B_{,r} \dot{\zeta} - B_{,\zeta} \dot{r} \right )
\right |_\Sigma, \\
\left . \frac{}{} \Omega_{,\rho} \dot{z} -
\Omega_{,z} \dot{\rho} \right |_{\Sigma} =
\left . - \frac{e^{4U}}{\rho} \left ( A_{,z} \dot{z} + A_{,\rho}
\dot{\rho} \right ) \right |_{\Sigma} = - \left .\frac{e^{4V}}{\alpha}
\left (  B_{,r} \dot{r} + B_{,\zeta} \dot{\zeta} \right ) \right |_{\Sigma}, 
\end{eqnarray*}
where the righthand sides are known. Thus, the matching conditions fix the
normal derivative of $\Omega$ on $\Sigma$ uniquely, but they only fix $\Omega$
on $\Sigma$ up to an arbitrary additive constant. As we shall see, this fact
will add some subtleties in the uniqueness proof below.

We are left with Ernst equations (\ref{EQ}) subject to boundary conditions both
at the fixed (and known) compact axially symmetric surface given by (\ref{BS})
and at infinity. The conditions at infinity ensure that the spacetime is
asymptotically flat (which is the mathematical translation of the isolation
of the body) and are given by
\begin{eqnarray}
U = -m R^{-1} + O(R^{-2}), \hspace{1cm} \Omega =-2 z J R^{-3} + O(R^{-3})
\label{coninf}
\end{eqnarray}
where $R \equiv \sqrt{\rho^2 + z^2}$ and $m$ and $J$ are the total mass and
angular momentum of the source, respectively. It is now evident
that the boundary conditions on $\Sigma$ constitute a Cauchy problem for
the {\it elliptic} system (\ref{EQ}), so that it is mathematically
not well-posed.
Thus, the physical problem is overdetermined and we should not expect existence
of solutions in all situations. This is the main problem
for proving the existence of the exterior solution given the interior,
and it is a matter of further investigation to find which restrictions
on $\left(V_I,g_I\right)$ will allow to overcome it.

Nevertheless, we can now address the proof of the uniqueness
of the exterior field, which is one of the purposes of this letter, by using
the theory of harmonic maps. Roughly speaking, harmonic maps
are applications between Riemannian manifolds that extremize
the so-called {\it energy or Dirichlet functional}.
More precisely, given an $n$-dimensional
Riemannian manifold (possibly with boundary) $(M,\overline{g})$, a
$k$-dimensional Riemannian manifold $(N, \gamma)$, and a $C^1$ map
$\Psi\, :\, (M,\overline{g}) \longrightarrow (N, \gamma)$,
we can construct the energy functional
$E(\Psi) \equiv \int_M \frac{1}{2} \left [ tr_M \Psi^{\star}
(\gamma) \right ]
\mbox{\boldmath $\eta$}$,
where {\boldmath$\eta$} is the volume form of $M$ and $tr_M
\Psi^{\star}(\gamma)$ is the trace in $M$ of the pull-back of the
metric tensor $\gamma$ of the target manifold $N$.
If we introduce local coordinates $\{x^{\mu}\}$ and $
\{ \Psi^a \}$ in $M$ and $N$ respectively, then $E(\Psi)$ becomes
\begin{eqnarray*}
E(\Psi) = \int_M \frac{1}{2} \overline{g}^{\mu\nu}(x)
\gamma_{ab}(\Psi(x))
\frac{\partial \Psi^a}{\partial  x^{\mu}}
\frac{\partial \Psi^b}{\partial x^{\nu}} \sqrt{\det
\overline{g}}\, dx^1 \wedge \cdots \wedge dx^n
\end{eqnarray*}
where $\mu,\nu$ run from $1$ to $n$ and $a,b$ run from $1$ to $k$.
Then, any $C^2$ map with finite $E\left (\Psi \right)$ is called
harmonic iff it is a critical point of $E$ with respect
to variations vanishing on the boundary of $M$ (whenever $\partial M
\neq \emptyset$). Thus, any harmonic map $\Psi$ satisfies the corresponding
Euler-Lagrange equations, which read (in local coordinates for $N$)
\begin{eqnarray}
\triangle \Psi^{c} + \Gamma^{c}_{ab}(\Psi(x))
 \langle d\Psi^a , d \Psi^b  \rangle = 0, \label{Harm}
\end{eqnarray}
where $\Gamma^{c}_{ab}$ are the Christoffel symbols
in $N$. It is well-known that Ernst equations (\ref{EQ}) 
can be written in this Euler-Lagrange form if the
domain manifold $M$ is chosen to be the part of the Euclidean 3-space
with extends from the two-dimensional
compact boundary $\tilde{\Sigma}\equiv \{\rho(\lambda), z(\lambda), \phi \}$
to infinity and the target manifold $N$
to be the two-dimensional manifold with metric
\begin{eqnarray}
ds_{N}^2 = 4 dU^2 + e^{-4U} d\Omega^2. \label{target}
\end{eqnarray}
(Notice that $\tilde{\Sigma}$ is in fact the image of a new imbedding of the
abstract $\Sigma$, in this case into $M$. Thus, even though there is a
one-to-one correspondence between $\Sigma$ and $\tilde{\Sigma}$,
they should not be confused as their corresponding first fundamental forms 
do not coincide). The solutions of (\ref{EQ}) we are interested in must
must have the asymptotic behaviour
(\ref{coninf}) which ensures that the corresponding  energy
functional is finite.

The proof we will give for the uniqueness of the exterior
gravitational field consists in two steps.
First, we will show that two solutions with the same Dirichlet
boundary conditions on $\tilde{\Sigma}$ must coincide. To that aim,
let us assume that
there are two harmonic maps $\Psi_0$ and $\Psi_1$ from $M$ to $N$ satisfying
$\left . U_0 \right |_{\tilde{\Sigma}} = \left. U_1 \right |_{\tilde{\Sigma}},
\hspace{3mm} \left . \Omega_0 \right |_{\tilde{\Sigma}} = \left . \Omega_1
\right |_{\tilde{\Sigma}}$. To proof that they coincide everywhere,
we will make use of a most remarkable identity due to Bunting \cite{Bu}.
The idea is to construct a smooth homotopic application
\begin{eqnarray}
\Psi: & & M \times \left [0,1 \right] \longrightarrow N \nonumber \\
& & \left ( x, u \right) \longrightarrow \Psi(x,u) \label{homot}
\end{eqnarray}
satisfying $\Psi(x,0) = \Psi_0$, $\Psi(x,1)= \Psi_1$ everywhere on $M$ and
such that $\gamma_{x_0}(u) \equiv \Psi(x_0,u)$ is an affinely parametrized
geodesic in $N$ for every $x_0 \in M$. In our case, $N$ is a
simply connected geodesically complete maximally symmetric plane with constant
negative curvature so that a well-known theorem (see e.g. \cite{Mth}) asserts
that any two points can be joined by a unique geodesic. Therefore the homotopy
(\ref{homot}) can always be constructed.
A remarkably simple calculation (see also \cite{Car3}) leads to
Bunting's identity
\begin{eqnarray}
\frac{1}{2} \nabla^{\mu} \nabla_{\mu} s^2 =
\int_{0}^{1} du \left \{\nabla^{\mu} s_i \nabla_{\mu} s^i - {R}_{ijkl}
s^i \nabla_{\mu} \Psi^{j} s^k \nabla^{\mu} \Psi^l \right \} \label{Bun}
\end{eqnarray}
where $R_{ijkl}$ is the Riemann tensor in $N$,
$s^{i}(x)$ is the tangent vector along the geodesic $\gamma_{x}$
and $s^2 (x)$ is its norm (being geodesic and affinely
parametrized $s^2$ does not depend on $u$). For $N$ with non-positive
curvature, the righthand side of this
identity is non-positive definite and it only vanishes when $s^i$
is covariantly constant. The lefthand side in (\ref{Bun}) is
a total divergence and its integral in $M$
can be transformed into a surface integral on the boundary
$\tilde{\Sigma}$ and the integral ``at infinity''.
The coincidence of 
$\Psi_0$ and $\Psi_1$ on $\tilde{\Sigma}$ implies
$ \left . s \right |_{\tilde{\Sigma}} = 0$ and consequently
the surface integral on 
$\tilde{\Sigma}$ vanishes. Regarding the integral ``at infinity'', the
asymptotic behaviour of $\Psi_0$ and $\Psi_1$ is of the form (\ref{coninf})
\begin{eqnarray*}
\Psi_D = \left ( -m_D R^{-1} + O\left(R^{-2} \right),
-2J_D\, z R^{-3} + O \left (R^{-3} \right) \right), \hspace{1cm} D=0,1 
\end{eqnarray*}
(we cannot assume a priori the same mass and angular momentum
for both solutions).
The geodesics of $N$ can be explicitly integrated so that the evaluation of
the geodesic distance between $\Psi_0(x)$ and $\Psi_1(x)$ at points $x$
lying on a spherical surface $R= R_0$ proves
that $s$ falls off much faster than $R_0^{-2}$ when $R_0$ tends to
infinity. Hence, the integral on the surface ``at infinity'' also vanishes and
we can conclude that $s^i$ is a constant vector field in $M$. Since  $s^{i}$
vanishes on $\tilde{\Sigma}$, $s^{i}$ must be zero everywhere and
$\Psi_{0}(x)= \Psi_1(x) \hspace{3mm} \forall
x \in M$ as we wanted to prove.

In the second part of the proof we must take into account
that, as shown above, the matching conditions fix
the value of $\Omega$ on $\Sigma$ (and thereby
on $\tilde{\Sigma}$) up to an
arbitrary additive constant {\it which is relevant} , because the
freedom in $\Omega$ has already been used to set the asymptotic behaviour
(\ref{coninf}). Therefore, we still need to show that this arbitrary
additive constant does not introduce a one-parameter family of solutions.
To that end, the  full set of boundary conditions (\ref{BC}) on $U$ and
$\Omega$ will now help. Let us assume there exist two solutions
$\Psi_{+} = \left (U_{+}, \Omega_{+} \right )$, and $\Psi_{-} = \left (
U_{-}, \Omega_{-} \right)$ satisfying
\begin{equation}
\left . U_{+}  \right |_{\tilde{\Sigma}} =
\left . U_{-} \right |_{\tilde{\Sigma}},
\hspace{2mm} \left . \nabla_{\mu} U_{+} \right |_{\tilde{\Sigma}}= 
\left . \nabla_{\mu} U_{-} \right |_{\tilde{\Sigma}},  \hspace{2mm}
\left . \Omega_{+}  \right |_{\tilde{\Sigma}} =
\left . \Omega_{-} \right |_{\tilde{\Sigma}}
+ C, \hspace{2mm} \left . \nabla_{\mu} \Omega_{+} \right |_{\tilde{\Sigma}}= 
\left . \nabla_{\mu} \Omega_{-} \right |_{\tilde{\Sigma}} \label{new}
\end{equation}
where $C$ is an arbitrary constant. A direct calculation shows that, given any
solution $\Psi =\left( U, \Omega \right)$ of the Ernst equations (\ref{EQ}),
the three-parameter one-form on $M$
\begin{eqnarray*}
\bm W\left (\Psi ;c_1,c_2,c_3 \right) \equiv 2\left(c_2+2c_3\Omega\right) dU +
\left[e^{-4U}\left(c_1+c_2\Omega+c_3\Omega^2\right)-c_3\right] d\Omega
\end{eqnarray*}
has vanishing divergence, (i.e. $\nabla^{\mu} W_{\mu}\left (\Psi \right)
\equiv 0$) and, consequently,
\begin{eqnarray}
\int_{\tilde{\Sigma}} W_{\mu} dS^{\mu} =
\int_{S_\infty} W_{\mu} dS^{\mu} = 8\pi c_2 m(\Psi) \label{Meq}
\end{eqnarray}
where $S_{\infty}$ stands for the surface ``at infinity'' and the last
equality follows immediately from the asymptotic flatness conditions
(\ref{coninf}). On the other hand, relations (\ref{new}) imply
\begin{eqnarray*}
\left. \bm{W} \left (\Psi_{+};c_1,c_2,c_3 \right)\right |_{\tilde{\Sigma}}=\left.
\bm{W}\left (\Psi_{-}; c_1+c_2C+c_3C^2, c_2+2c_3C, c_3 \right)
\right |_{\tilde{\Sigma}} \, .
\end{eqnarray*}
The combination of this expression with the identity (\ref{Meq}) produces
\begin{eqnarray*}
8 \pi c_{2} m(\Psi_{+}) = 8 \pi \left (c_2 + 2c_3C \right) m(\Psi_{-})
\end{eqnarray*}
and, given that $c_2$ and $c_3$ are arbitrary constants, it necessarily follows
\begin{eqnarray*}
m\left (\Psi_{+} \right)  = m \left (\Psi_{-} \right),
\hspace{1cm} m(\Psi_{-}) C  =0.
\end{eqnarray*}
Since for physically well-behaved solutions the total mass cannot
vanish, $C$ must be zero and the
uniqueness of the exterior solution generated by a given interior distribution
of matter in stationary and axially symmetric rotation is completely proven.

\end{document}